\documentclass[3p,final]{elsarticle}

\usepackage{amssymb}
 \usepackage{amsthm}
\usepackage{amsmath}
\usepackage{microtype}
\usepackage{graphicx}
\usepackage{algpseudocode}
\usepackage{algorithm}
\usepackage{complexity}
\usepackage{hyperref}
\usepackage{graphicx}

\journal{arXiv}

\begin{document}

\begin{frontmatter}


\title{eXpOS: A Simple Pedagogical Operating System for Undergraduate Instruction}
\author{K.~Murali Krishnan \fnref{fn1}}
\ead{kmurali@nitc.ac.in}
\address{National Institute of Technology Calicut, India.}
\fntext[fn1]{Corresponding on behalf of the project team, whose 
members are listed in Section~\ref{authors}.}

\title{}


\author{}

\address{}

\begin{abstract}
An operating system project suitable for undergraduate computing/electrical sciences students is presented.  The project can be used as a course project in a one semester course, or as a self-study project for motivated students.   The course is organized such that a student with a basic background in programming and computer organization can follow the implementation road map available online, and build the OS from scratch on her personal machine/laptop, with minimal instructional supervision.  The student is provided with a simulated abstract machine, an application interface specification, specification and design of the OS, and a step by step project implementation road map.  The functionalities of the OS include multitasking, virtual memory, semaphores, shared memory, an elementary file system, interrupt driven disk and console I/O, and a limited multi-user support.  The final stage of the project involves porting the OS to a two-core machine.  
An independent one semester compiler design project, where the student builds a compiler for a tiny object oriented programming language that generates target code that can be loaded and executed by the OS is also briefly discussed.    
\end{abstract}


\end{frontmatter}



\section{Introduction}

Programming projects prescribed to students in undergraduate OS courses fall into two main categories. The first category consists of assignments that require the use of the system call interface of an existing operating system, but not requiring the students to write kernel code.  Asking the student to program a shell, implement a solution for the ``dining philosopher’s problem” using semaphores and shared memory system calls or write her own heap management routines,  are exercises that fall into this category. These exercises enhance the student's understanding of the OS functionality, but are not intended to give serious insights into the design of an OS.

Our interest is in the second category of kernel projects - programming assignments that ask the student to code a part or whole of the OS kernel. An ideal kernel project would ask a student to build an OS kernel completely from scratch.  However, writing kernel code on real hardware to achieve non-trivial outcomes is too complex a task for an undergraduate student, particularly considering the fact that she will be crediting several other courses concurrently.   Hence one among the two compromising pedagogical choices below needs to be made.   

The first route is to ask the student to modify/enhance certain parts of a given OS kernel code base, implemented on some real/simulated hardware machine. Minix~\cite{Minix}, XINU~\cite{Xinu}, PintOS~\cite{Pintos}, Topsy~\cite{Topsy}, TempOS~\cite{Tempos}, GeekOS~\cite{Geekos}, Nachos~\cite{Nachos}, OS/161~\cite{OS161}, OSP~\cite{OSP} and Xv6~\cite{Xv6} provide such environments.   The student is asked to re-implement/modify certain central components of the kernel, such as scheduler, memory manager, or add a new file system, as decided by the course instructor. Projects of this kind can be quite intimidating. For instance, Pinto et. al.~\cite{Tempos} reports that 36\% of the students who enrolled in an OS course dropped out on the first day when the course project (TempOS) was announced.  

The second route is to ask the student to code the complete kernel from scratch without any given code base, on some  “pedagogical hardware'' designed to make the task manageable as a course project. In the classic Nand2Tetris project~\cite{Nisan},  the student starts with NAND gate and D flip flop as primitives to build a small machine,  and then goes on to add a whole software suite consisting of an assembler, a compiler and an OS on top of it. Crowley~\cite{Crowley} outlines a project where the student conceptualizes an OS on the hypothetical CRA-1 architecture. Such projects are more self-learnable as they start from scratch, and provide the student a complete picture of the system she builds. The compromise involved is that the student works in a custom educational environment, rather than on real hardware and professional code bases.  The OS project described here belongs to this category. The student works on a hypothetical machine that treats strings as primitive elements (allowing an arbitrary string to be stored in a memory word).  She is asked to write the complete OS kernel for this machine in an enriched assembly language, which can be compiled and run using an assembly language interpreter for the machine.

In comparison with the Nand2Tetris project, the present project limits its scope to operating systems.  However, the project makes the student build a more ambitious OS, that supports cooperative multitasking,  virtual memory management, interrupt based disk and I/O handling, a rudimentary file system, limited support for multiple users and a very basic two-core implementation.   The project is similar in philosophy, sequencing of topics and complexity to~\cite{Crowley}, though technical details and pedagogical considerations differ.  The abstract machine for which the student writes the OS is specifically designed to make kernel programming simple.  Students without a strong programming background can easily get started with the project and pick the skills needed as they go on.  The project documentation contains a detailed implementation road map that enables the student to self-learn all the background required for the project and complete the project in a step by step manner with minimal instructor intervention.   The complete kernel requires the student to write around 3000 lines of (enriched) assembly language code.  

eXpOS is an open source project, developed as a sequence of student projects at National Institute of Technology Calicut, India.  The project documentation, simulator and other support software can be downloaded from the Github repository of the project \href{https://exposnitc.github.io}{exposnitc.github.io}.  The user is required to have a Linux system to run the hardware simulator.      

An independent one semester pedagogical project which asks the student to write a compiler (as well as associated linking and heap management routines) for a tiny object oriented programming language (called OExpL), so as to generate target assembly code that is compatible with the eXpOS loader, is briefly described in the end. 
(Documentation, simulator and other support software for the compiler project can be downloaded from the Github repository \href{https://silcnitc.github.io}{silcnitc.github.io}.) Neither the compiler project nor the OS project assumes the other one as a pre-requisite or a co-requisite.

\section{Pedagogical Considerations}

Our goal was to design an OS project suitable for CS/Electrical undergraduate students with basic background in programming and computer organization, that helps the student to comprehend a “sufficiently non-trivial” operating system kernel as a whole functional unit, understand how it interacts with (and what it expects from) the lower hardware layer as well as the upper application layer, and appreciate how the complex system can be broken down to component subsystems that can be joined together to form the whole.  

The design of the system has been guided by experience and feedback from several faculty members (including the author) who have conducted OS laboratory experiments and projects with undergraduates, both with Minix~\cite{Minix}  and Nachos~\cite{Nachos} systems at NIT Calicut, India,  between 2003 and 2013.   Hence the pedagogical considerations presented here are invariably bound to teaching-learning conditions and prior training of both students and teachers in the region.  

A large number of students enrolling into the course had never written a program exceeding 500 lines of code, and lacked the proficiency needed to read and understand the Nachos/Minix code base.  The class strength was too large to permit individual attention to each student on a regular basis. Hence, we decided to design a new OS project where the student can proceed on her own, start with baby steps, gain confidence,  and gradually proceed to bigger steps, building one’s own code base.  This led us to place the  following requirements on the proposed system:  

\begin{itemize}
 \item The student must be able do the project with minimal instructor intervention. 
 \item She should build the kernel following a road map, through a large number of small stages, each stage making incremental additions to the previous stage code, with no externally supplied kernel code to be included at any stage.
 \item Each stage of the road map must precisely define the learning objectives of the stage, and provide sufficient information needed for completing the particular stage.
 \item After the initial stages, the student should have a feeling of self belief that she has already completed a bare minimum OS, and it is just a matter of time before she develops the system into a fully functional OS.  That is, the road map must outline a spiral model of development.  
 \item Each student may work through the road map at her own pace and reach up to where she is able to (at the end of the course),  and achieve commensurate learning outcomes.  (From the instructor's perspective, the student may be graded based on the stage she has completed satisfactorily.)
\end{itemize}

OS programming cannot be simplified unless the underlying hardware is easy to comprehend and program. Similarly, the interface of the OS to the application must be kept simple, minimal and comprehensible to an undergraduate student,  so that writing an OS that meets the functional requirements is both intellectually and practically manageable. These considerations  led to the following requirement specification.

The student needs to be provided with the following documentation:  

\begin{itemize}
 \item Hardware: A simple instruction set architecture abstraction, that makes it is possible for the student to write within a few hours, a bootstrap loader that prints “hello world.''  The machine needs to support multiprogramming, virtual memory, interrupt based  disk and I/O etc.  Sufficient hardware abstractions must be provided so that intricacies of disk and input/output systems do not hamper the progress of the project.  
 
 \item Application Interface:  A simple interface between the OS and the application, that allows the student to figure out in a few hours how to load and execute an application from the disk into memory and execute it.  At the same time, the system calls must be powerful enough to create and modify files,  support concurrent processes, allow processes to share memory and files with proper concurrency control primitives and allow users to set file access permissions.
 
 \item OS design:  The design of a kernel that implements the application interface on top of the the hardware.  Outline of the algorithms and kernel data structures also needs to be documented and supplied.    
 
 \item Road map:  An implementation road map that splits the kernel writing process into a spiral step-by-step process so that  the student can follow the steps of the road map and implement the project with minimal instructional supervision.  
\end{itemize}

Once these requirements were formulated, the associated tasks included implementing a hardware simulator, compilers/assemblers to support programming the hardware, and support tools for transferring files between the machine's disk and the external environment.  

\section{Description of the Learning System}

This section provides some technical details of the learning system.  Readers not interested in details can skip to the next section without loss of continuity.  

\subsection
{\href{http://exposnitc.github.io/arch_spec.html}{Hardware Specification}}
The hardware consists of a machine,  called “Experimental String Machine (XSM)”, which is a simple paged architecture that treats strings as primary data objects.  The machine hardware consists of a CPU, disk, memory console I/O and a timer.   When the simulator is started, the firmware loads the contents of the first disk block (where the student must put her OS startup code) to the second memory page and transfers execution to the loaded code.   The student must code the kernel in an enriched assembly language (called the SPL language in the project documentation) and load the compiled code to the disk externally.  A support software called “XFS-Interface” is provided to load programs from the host machine to the XSM disk.  XSM hardware supports interrupt driven disk and console I/O.  

\subsection
{\href{http://exposnitc.github.io/abi.html}{Application Interface Specification}}
The application interface stipulates i) how applications can invoke system calls (including details of call stack parameter passing conventions) ii) how the the OS expects the application to partition the logical address space into code, heap and stack etc and iii) The specification of the arguments to be passed to each system call and the interrupt number to which each system call is mapped into.

\noindent \textbf{Note:} The XSM machine simulator provided to the student is an assembly language interpreter.  Hence, the application binary interface specification, strictly speaking, is an “application assembly interface”.

\subsection
{\href{http://exposnitc.github.io/os_design.html}{OS Specification and Design}}
Documentation providing a detailed specification and design of the OS kernel and an outline of the data structures and algorithms for each system call, module, exception, interrupt and OS module, which the student is expected to read and grasp as and when needed, is given (See Figure 1). Some implementation details (such as the static memory regions where each OS data structure and routine must be pre-loaded from the disk during bootstrap) are fixed by the design.  These specifications are also part of the documentation.   

\subsection{Support tools} 
\begin{itemize}
 \item \href{http://exposnitc.github.io/support_tools-files/xfs-interface.html}{XFS interface:}  
 This is a software for transferring data files and executable files between the host (Linux/Unix) system and the XSM machine’s disk.    
 \item \href{http://exposnitc.github.io/support_tools-files/expl.html}{ExpL Compiler:}  
 The application interface of the OS is too limited to interface to C programs.  Hence a cross-compiler for a custom high level language called ExpL using which application programs can be written and compiled to the target application interface is provided. The student writes and compiles applications from the host system and transfers the executable code to the XSM machine’s disk using the XFS interface tool.  ExpL does not support kernel programming.    
 \item \href{http://exposnitc.github.io/support_tools-files/spl.html}{SPL Compiler:}  
 A compiler for a programming language called SPL (essentially an enriched assembler for XSM machine) in which students can write OS kernel code.  The student writes and compiles kernel code from the host system and transfers the executable code to the XSM machine’s disk, using the XFS interface tool.   
 \item \href{http://exposnitc.github.io/support_tools-files/xsm-simulator.html}{Hardware Simulator and Debugger:}  
 Hardware simulator for the XSM machine with a built in kernel code debugger, providing an interface similar to gdb.
\end{itemize}
To give the reader a feel of the programming environment given to the student, we provide below an SPL code snippet that runs a counter up to 10 in machine register R0.  If the simulator is run in debug mode, then the debugger is entered whenever the breakpoint is executed (the instruction is ignored otherwise).  
\begin{verbatim}
           alias counter R0;
           counter = 0;
           while(counter <= 10) do
                 if (counter%2 != 0)  breakpoint; endif;
                 counter = counter + 1;
           endwhile; 
\end{verbatim}

\subsection{\href{http://exposnitc.github.io/Roadmap.html}{Road Map}} 
The twenty eight stage road map guides the student through the building of the OS following a spiral model of program development.    We sketch below the flow of kernel development laid down by the road map. 

\begin{figure}\label{fig:Design}
\begin{center}
\includegraphics[width=\textwidth]{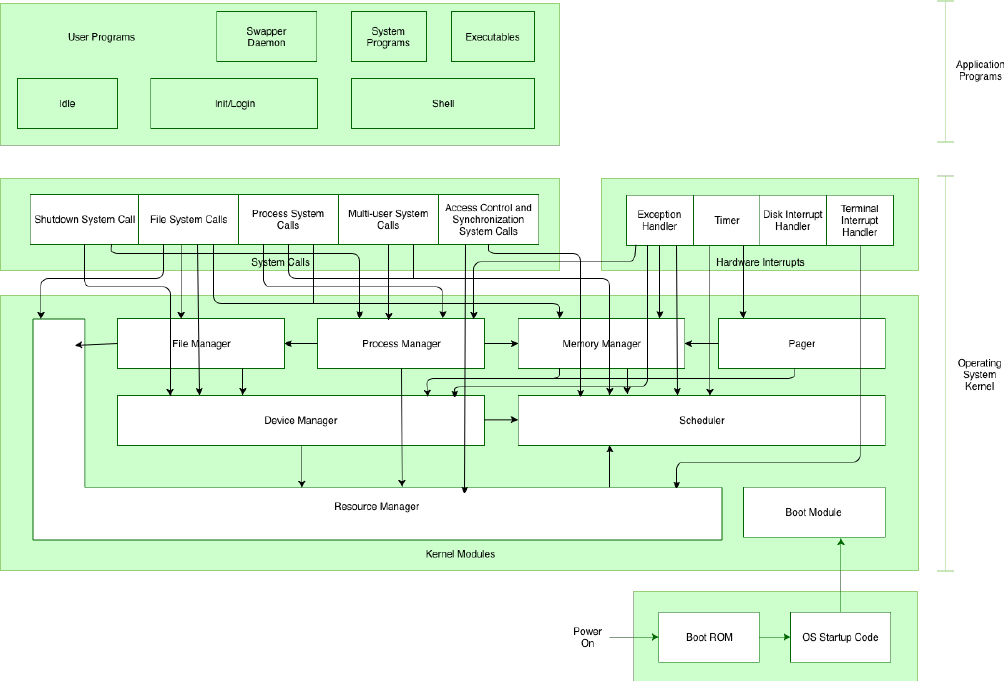}
\caption[]{Modular Design of eXpOS}
\label{Serie}
\end{center}
\end{figure}

The first twelve stages are preliminary stages (baby steps) starting with stage 0 (installation instructions).  Each stage takes no more than 2-4 hours to complete.  The first five stages are tutorials on the file system format, use of the tools for host-simulator file transfer, use of the assembler and debugger etc.  The student writes a simple bootstrap loader that prints “Hello World” by the end of Stage 3. The next few stages help the student to understand the paging mechanism of the machine, call stack conventions between the application and system calls (traps) and terminal output.  By the sixth stage, the student will complete an OS that on boot-up, loads a single application (that can execute instructions and print to the console).  The next few stages take the student through timer interrupt handling and task switching.  By the end of Stage 12, the student will have implemented a tiny OS kernel that can load a predetermined set of applications from the disk at boot-time and run them to completion concurrently, but only permitting console output.  Context switch can happen only on timer interrupt at this stage.  

The next seven stages (called intermediate stages) begin with the management of kernel context, necessary to allow context switching when an application is executing in kernel mode, inside a system call.  This is followed by stages that explain interrupt based console/disk input and blocking of processes while waiting for interrupts.  The student completes most of the interrupt/exception and context management routines during the intermediate stages.   By Stage 18, the student builds an OS that upon boot, runs a primitive shell program that asks for the name of an application, loads and executes the application and halts.  Stage 19 adds exception handler code for handling page faults so that only code pages that are needed are loaded for execution when the application runs.  The student completes an initial version of the memory manager of the OS that does all the memory management functions except disk swap, by Stage 19.   (Swapping of processes to disk is delayed till Stage 27).  At this point, the student completes implementation of the first system call - Exec.  Each intermediate stage is designed to be doable in less than six hours.  However good skill in using the debugger is necessary to complete the work within this schedule, as bugs in concurrent code can be hard to find and may cause delays.  

 In Stage 20, the student completes Fork and Exit system calls;  Stage 21 adds Signal-Wait primitives and Stage 22 adds Semaphores.  In Stage 21, the student implements a shell and at this point, she has an operating system that repeatedly asks the user to enter a command (name of an application) that it will execute.  The shared memory facility provided by the OS is such that the parent and the child processes share the parent’s heap upon execution of the fork system call.   Applications can control access to the shared heap using semaphores.  During Stages 23-25,  the student adds file system calls to the kernel (Create, Delete, Open, Close, Read, Write and Seek).  The OS permits only single word read/write operations on a file from the current seek position.   A few recently accessed blocks are loaded on demand into a fixed-size buffer cache and a process sleeps if the block needs to be loaded from the disk to the cache.    Stage 25 is a good point for an average student to reach at the end of the course.  

Stage 26 adds limited multi-user support to the OS, with only one user allowed in the system at a time (there is only one console for the system).  Basic user management system calls (Login, Logout, and system calls to add/remove users) are implemented in this stage.  The student also modifies file write and delete system calls to ensure that only the owner of a file can modify/delete data files unless permission is given (users have read access to all files).  Stage 27 adds virtual memory management functions to the OS.  At this point, the system will be able to swap out processes and run the maximum allowable number of processes concurrently (determined by the number of entries in the task structure, which is set to 16).        

A final advanced stage added subsequently (Stage 28) asks the student to port the OS to run on a two-core SMP machine (called NEXSM).  In this stage, the student learns how to use the TSL instruction to manage parallelism.  

\section{Classroom Experience}           

Third year operating systems undergraduate laboratory courses were conducted in Fall 2018 using the eXpOS system at National Institute of Technology Calicut (NIT Calicut) by and Indian Institute of Technology Palakkad (IIT Palakkad), India by different teams of instructors.  The students in both the institutions were admitted to the respective institutions through the all India admission procedure and had comparable backgrounds.    However, the course at NIT Calicut was offered as an elective where only interested students joined the course,  whereas in the IIT Palakkad curriculum, OS lab was a core course.  Feedback collected -from 42 respondents at NIT Calicut and 19 respondents at IIT Palakkad is analyzed below.  The course was offered again as an elective in NIT Calicut in Fall 2019 with 58 respondents. 
 
\subsection{Student Performance}
Almost every respondent reported that they could complete Stage 20 (process management)  in both the courses (100\% at NIT Calicut in both 2018 and 2019,   95\% at IIT Palakkad).    76\% of (respectively 78\%) of the students in NIT Calicut in 2018 (respectively, 2019) reported completion up to Stage 25 (file system implementation),  whereas this percentage was 48\% at IIT Palakkad.  The data shows that the course is more effective when offered as an elective to motivated students and that the average and weak student are likely to struggle in the course beyond Stage 20.   33\% among those who completed Stage 25  at IIT Palakkad (respectively 47\%  and 28\% at NIT Calicut in 2018 and 2019 respectively) went on to complete up to 27 stages of the project (user management and virtual memory management).   In general, we expect nearly a third of the students reaching Stage 25 to reach Stage 27 or further).  9\% of the students at NIT Calicut went on to complete all stages of the project including the advanced multi-core implementation stage (Stage 28) in 2019.  (This stage was not designed in 2018).

\subsection{Student Preparation}   50\% (respectively, 42\%) of the students at NIT Calicut (respectively, IIT Palakkad) in 2018 noted that their background in computer organization was inadequate.   Interaction with students brought out the following issue.  Computer organization courses are typically based on the MIPS architecture and lay more focus on design issues, and tend to place less stress on page table address translation as well as the semantics of instructions such as push, pop, call, ret and interrupt/exceptions that manipulate the stack, that are needed for the project.  In fall 2019, a few initial lectures discussing the required background in computer organization were conducted and this reduced the percentage of students reporting background deficiency in computer organization to 35\%.

\subsection{Road map Effectiveness}   88\% (respectively, 93\%) of the respondents from NIT Calicut in 2018 (respectively, 2019) responded that 90\% of the project can be done completely based on the road map of and associated documentation of the project.   However, only 53\% of the respondents at IIT Palakkad gave the same response.  Discussions with students indicated that average and weak students had difficulty in understanding certain parts of the documentation.  We have added some detailed explanations in certain sections of the documentation based on the feedback.  

\subsection{Learning Objectives}  88\% (respectively, 93\%) of the respondents at NIT Calicut in 2018 (respectively, 2019) indicated that doing the lab gave a clear advantage in understanding in OS theory.   68\% of the respondents at IIT Palakkad reported so. One of the facts that seemed to have contributed to the difference, apart from the core vs elective issue is that the OS theory course offered at IIT Palakkad was for a small class of around 30 students, and followed the book~\cite{Crowley}, which focused on kernel design issues.  The course offered at NIT Calicut was credited by more than 150 students, followed the book~\cite{Silberschatz}, and was more concepts oriented.  As the former text covers many of the design issues encountered during the project, the enforcement that the laboratory gave to the understanding of theory was less dramatic. 

The feedback suggests the following for an effective conduct of the course.  

\begin{itemize}
 \item The project is best suited to be offered as a one semester elective course for motivated undergraduates.  In such cases, around 75\% of the students may be expected to complete the core part of the project consisting of process management, basic memory management (except swapping), process control  and file system implementation effectively.  
 \item A few lectures refreshing the relevant topics in computer organization such as hardware paging and address translation, handling of interrupts and exceptions etc. at the beginning of the course will be effective.
 \item Most students can be expected to do most of the project following the road map and associated documentation, without requiring additional inputs from the instructors.   
\end{itemize}

Finally, we observed that students who initially had a fear for large code, were free from the fear at the end of the project.  
  
We add a few comments that may help instructors in running the course.  One issue with such programming projects is that students may collaborate with each other and a few of them may engage in copying code from other students.   An evaluation strategy that has been effective, is to conduct examinations of the following format.  The student would be given a small change in the specification of a system call.  The change is to be designed so that several parts of the OS would require very minor edits to make the change, but the overall coding required is small.  The student is first asked to submit in 30 minutes a design write up that specifies which routines of the kernel would require change and what changes are to be made.  They can take 3-4 hours from then to submit the modified code.  The instructors check whether the design is correct and modifications to the code were in conformation with the design, before running test cases.  We laid more stress on the design rather than the working of the code.  

Regarding the conduct of the course, we permitted students to discuss the project with other students, but insisted that they do the code themselves.  This is with the belief that encouraging discussions will boost  the learning process, particularly in a self learning environment.  When the course was run in 2019, a set of senior students who volunteered to mentor their juniors in doing the project were identified and each student was assigned a mentor who would help them with support and guidance in doing the project themselves.  This too proved very effective, particularly in helping students with debugging.  It should however be noted, that teaching-learning methods that prove effective in one society may need suitable alterations to work well elsewhere. 

\section{OExpL:  A Companion Compiler Writing Project} 

Graduate courses in compiler design tend to move quickly from the basics to intermediate code representation, frameworks such as LLVM~\cite{LLVM} and proceed to code optimization and back-end issues.   The goal of the project presented here is to cover material that prepares an undergraduate student for future graduate courses.  Apart from standard lexical/syntax analysis, type/scope resolution and code generation issues, the student comes across topics such as linking, heap management and run time binding of polymorphic functions. (Unfortunately, many undergraduate CS programs pay little attention to these topics).  The student is given the specification of a toy object oriented programming language called OExpL.  The project requires her to implement a compiler and linker that does all the work necessary to build a target file that is recognizable to the eXpOS loader.   Hence, a student who completes both the projects will be able to obtain a clear picture of how an application passes through compilation, linking, loading and execution steps. However, the project is designed to be accessible to students who have not done the OS project.  

The methodology of the project is similar to the OS project.  A simulator for the XSM machine with a primitive OS implementation that can load and execute application programs, is given to the student.  Specification of the target application interface which the target code must satisfy is also also given.   A detailed implementation road map that guides the student in building the compiler, following a spiral model of program development, is provided.   

A brief description of the road map, which is divided into eight stages, is given below.  The first four stages are straightforward. The initial stage (Stage 0) contains simulator installation instructions and a tutorial on the application interface to the OS, with some practice exercises.  This stage also contains elaborate hands-on tutorials on Lex, Yacc and interfacing Lex with Yacc~\cite{Bison}.  
The stage equips the student with all the software tools required to do the project (and takes around 2 weeks of 5-10 hours of work).    
The rest of the project is about hand-crafting the compiler, working through the stages described next, writing a total of around 2500 lines of Lex, Yacc and C code.  

Stage 1 asks the student to build a small compiler for arithmetic expressions.  Stage 2 extends the language by adding support for a predefined set of static variables and assignments.   Stage 3 adds \texttt{if-then-else}, \texttt{while-do}, \texttt{break} and \texttt{continue} constructs.  Stage 4 adds arrays and string type variables and allows the user to name variables in the program.  Symbol table management and type checking are introduced in this stage.  Starting from Stage 3,  the student runs a first pass on the source program to generate an abstract syntax tree (using Lex-Yacc syntax directed translations scheme), runs a second pass that traverses the tree to generate target assembly code with labels,  and a final two-pass linking step to translate labels to addresses.  Students typically need 4-5 weeks (working 5-10 hours per week) to complete the project up to Stage 4.   Every student who has credited the course, has completed Stage 4.  

Stage 5 adds functions to the source language.  Automatic variables, scope issues as well as run time stack management issues are introduced here.  The next stage (Stage 6) extends the language to support user defined types (structures).  When a variable of a user defined type is declared,  the compiler is required to allocate only static/automatic memory for storing a reference variable.  The programmer has to allocate memory in the heap using the Alloc() function and assign it to the reference variable.  Hence, in Stage 6, the student is also asked to build a dynamic memory allocation library that implements Alloc(), Free() etc. These two stages together require around  4-5 weeks work (5-10 hours per week) to complete.   Stage 6 is a satisfactory point for an average student to reach at the end of the course.  

The two final stages of the project extends the language to support object oriented features.  In Stage 7, classes (that can hold both data objects and methods) are added to the language.  The final stage (Stage 8) adds support for single inheritance, implemented using a statically allocated virtual function table.  These two stages typically require another 2-3 weeks of work.   

We conclude the section with a few comments for instructors.   The project requires fairly strong programming background from the student.  An elective course based on the project has been offered for junior undergraduate  students at NIT Calicut in 2017, 2018 and 2019.    Typically 15-30 students enrolled for the course each year and most of the participants were those who did the OS lab (or some other heavy programming courses) earlier.  The project uses front end tools (Lex and Yacc) but does not go into back-end tools or frameworks such as LLVM.   The student is asked to implement code to traverse the abstract syntax tree recursively and generate code statement by statement.  No code/register optimizations are attempted and the back-end functionality is limited to generating semantically correct code.  On the other hand, the project is front-end intensive.  Only a skeleton of the grammar for the source language is given to the student, asking her to fill in the details.  As with the OS project, no code base is given to the student and she has to implement the data structures for class table, type table, symbol table, abstract syntax tree etc.  (We permitted students to use standard C/C++ libraries).  She also has to write the dynamic memory management routines. 

The project is available online as an open source Github repository, 
accessible from \href{https://silcnitc.github.io}{silcnitc.github.io}.  
As with the OS project, the hardware simulator requires a Linux machine to run.

\section{Project Team}\label{authors}
The eXpOS operating system project and the OExpL compiler project were developed by several teams of students at NIT Calicut, through a sequence of undergraduate projects during the period 2012-2018, under the supervision of this author, who is corresponding the paper on behalf of the team.  The team members are:
Ajeet Kumar,  Akhil S., Albin Suresh, Arun Joseph, Arun Rajan, Aswathy T. Revi, Avinash,  C.H. Vishal,  Deepak Goyal, Dileep Mathew Thomas, Gautham R. Warrier, Glen Martin, Govind R., Jampala Ritesh, Jaini Phani Koushik, Jeril K. George, Jithesh Kumar O. V., K. Dinesh,  Kandhala Naveena, Karthika Aravind, Kruthika Suresh Ved, Madisetty Jayaprakash, Mathew Kumpalamthanam, N. Ruthvik, Nachiappan V., Naseem Iqbal, Navaneeth Kishore, Nikhil Sojan, Nitish Kumar, Nunnaguppala Surya Harsha, Peeyush Singh, Ramnath Jayachandran, Rohith Vishnumolakala, Sathyam Doraswamy,  Shajahan Fariz, Sikha V. Manoj, Shamil C. M., Sreeraj S., Subin Puleri, Sonia V. Mathew, Subhisha, Sumedha Birajdar, Sumesh B., Thallam Sai Sree Datta, Vishnu Priya Matha, Vivek Anand T. Kallampally and Yogesh Mishra.  We also acknowledge the support from the faculty members Saidalavi Kalady, Saleena N. and Vineeth Paleri of NIT Calicut for technical support and Jasine Babu, Sandeep Chandran and Vivek Chaturvedi of IIT Palakkad who conducted the OS course at IIT Palakkad and provided us with feedback.

\end{document}